\documentclass[preprint,12pt]{elsarticle}

\usepackage{amssymb}

\usepackage{lineno}
\usepackage{color}

\journal{Nucl. Instr. and Meth. in Phys. Research Sect. A}

\begin{document}

\begin{frontmatter}



\title{A data acquisition system based on ROOT and waveform digital technology for Photo-Neutron Source}


\author[sinap]{L.X. Liu}
\ead{liulongxiang@sinap.ac.cn}

\author[sinap]{H.W. Wang\corref{cor1}}
\ead{wanghongwei@sinap.ac.cn}

\author[sinap]{Y.G. Ma}
\author[sinap]{X.G. Cao}
\author[sinap]{X.Z. Cai}
\author[sinap]{J.G. Chen}
\author[sinap]{G.L. Zhang}

\author[sinap]{J.L. Han\corref{cor1}}
\ead{hanjianlong@sinap.ac.cn}

\author[sinap]{G.Q. Zhang}
\author[sinap]{J.F. Hu}
\author[sinap,cas]{X.H. Wang}
\author[sinap,cas,shu]{H.J. Fu}

\cortext[cor1]{Corresponding author}

\address[sinap]{Shanghai Institute of Applied Physics, Chinese Academy of Sciences, Shanghai 201800, China}
\address[cas]{University of Chinese Academy of Science, Beijing 100080, China}
\address[shu]{School of Materials Science and Engineering, Shanghai University, Shanghai 200444, China}

\begin{abstract}
The data acquisition system is based on ROOT and waveform digital technology, including neutron detector,
waveform digitizer, PCI card, optical fiber, computer, reaction target device, stepper motor,
data acquisition software and control target software.
It achieves to acquire and record the waveform information of signal measured by the detector
using a waveform digitizer.
The specific target position is changed by the stepper motor which
is remotely controlled by the data acquisition software and control target software.
It is implemented by the exchange of information between the data acquisition software and
the control target software.
The system realizes to automatically open files and change targets at fixed intervals.
It is capable of data compression by removing the data those are not signals,
and automatic alarm when the beam is lost.
\end{abstract}


\end{frontmatter}

\section{Introduction}
\label{introduction}
The neutron energy is measured accurately by the time-of-flight (TOF) technique at the Photo-Neutron Source (PNS)
~\cite{jiang,lin1,lin2,ji,peng,wu}.
The PNS was constructed for the acquisition of nuclear data,
such as neutron total cross section and capture cross section, etc..
The measurement of neutron total cross section requires the data of sample target, blank target,
calibration target, and background target.
In order to avoid the influence of accelerator stability in long time operation,
it needs to change different targets in sequence at fixed intervals.
At the same time, it also needs to record the target position and the file number.
Because of the long experiment, the target needs to be changed frequently
that makes the manual operation hard and time consuming.
Therefore, automation becomes an inevitable choice.

With the improvement of computer and data storage capacity,
waveform digitizers  have developed rapidly in recent years~\cite{korn,laptev}.
It directly receives the pulse waveform of the detector without the amplifier, logic circuit and so on,
that makes electronics easier.
The time, amplitude, area and other information can be extracted directly from the waveform sampling points
and trigger timing tag.
In addition to these advantages, it also raises the question of huge data.
In a word, it is the most advanced data acquisition and recording mode at present.

In order to meet the special requirements of neutron total cross section, the data acquisition system
 is required to achieve multi-purpose, multi-parameter data acquisition function.
 Therefore, it is developed based on ROOT and waveform digital technology,
 and combines the reaction target device and control target software.

\section{Frame of the system}
 \begin{figure}
  \centering
  \includegraphics[width=14cm]{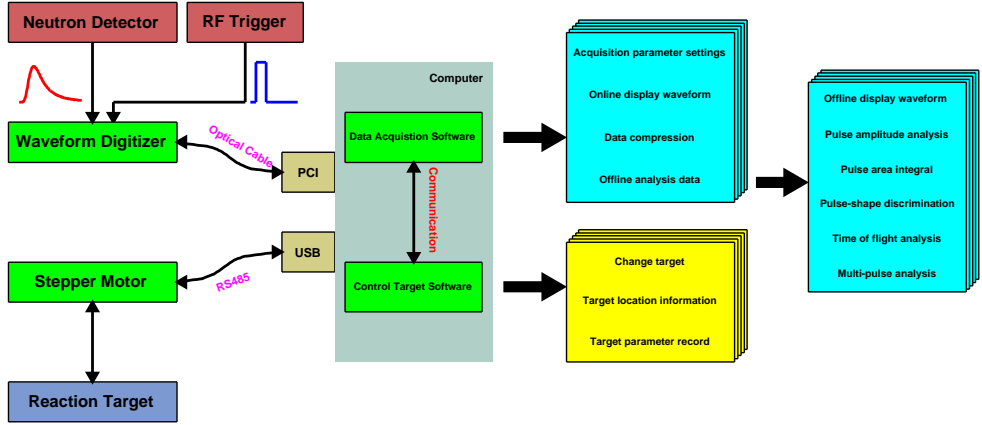}\\
  \caption{The architecture of data acquisition system for the PNS.}\label{frame}
\end{figure}

The data acquisition system consists of four main parts, which are waveform digitizer,
data acquisition software, stepper motor and control target software, as shown in Fig.~\ref{frame}.
The signals of neutron detector and high frequency trigger of accelerator are directly connected to the waveform digitizer,
which is controlled by the data acquisition software.
The collected data is transferred to the computer via optical fiber for storage.
the data acquisition software communicates with the control target software via a computer internal port.
The data acquisition software controls the stepper motor by exchanging information with the control target software.
The data obtained from the data acquisition software can be displayed online or analyzed offline.
\section{Components of the system}
\subsection{Waveform digitizer DT5720}

The waveform digitizer used in the data acquisition system is DT5720,
which is a four-channel, 12 bit device produced by CAEN, Italy.
Its sampling frequency is 250 MHz, i.e., the time of each sample is 4 ns,
and there is a 2 Vpp (peak to peak) signal dynamic range.
The DC offset of each channel is adjustable using a 16 bit DAC in the $\pm$1 V range.
A circular memory buffer is used to store the data stream continuously.
When the trigger occurred, an additional N samples after the trigger was stored by the FPGA
and the buffer was locked.
Then, it could be read via optical link or USB.
A new buffer was written continuously without dead time.
The transmission rate of optical link could reach 80MB/s.

\subsection{Data acquisition software}

\begin{figure}
  \centering
  \includegraphics[width=14cm]{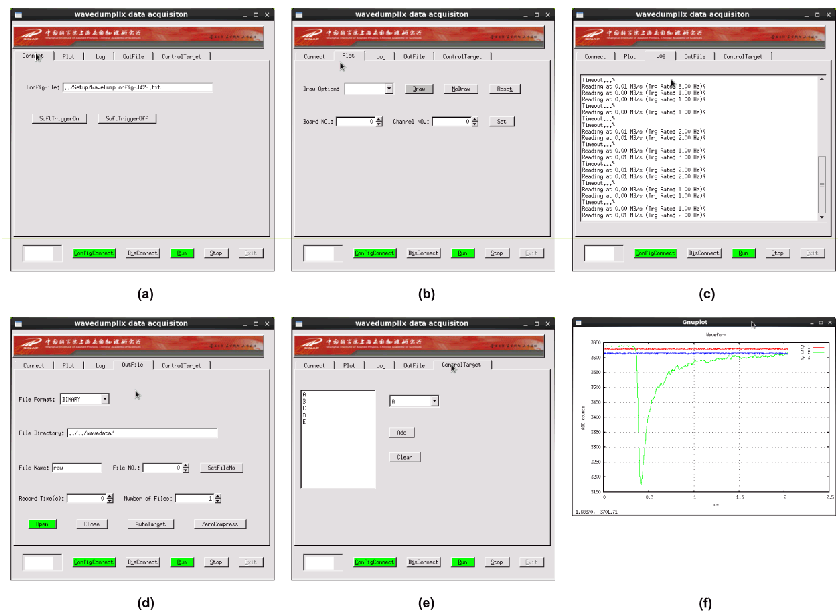}\\
  \caption{The operation interface and data online display of data acquisition software. (a) Connect module;
  (b) Plot module; (c) Log module;(d) OutFile module; (e) ControlTarget module; (f) Gnuplot display.}\label{wdas}
\end{figure}

 The data acquisition software is developed based on ROOT, Gnuplot,
 and hardware driver of digitizer.
 The software interface consists of two parts:
 the upper part is a common area, and the operating interface of the five modules can be switched freely,
 which are Connect, Plot, Log, OutFile and ControlTarget;
the bottom part is the interface of the public module, as shown in Fig.~\ref{wdas}.

The Connect module is used to set the storage path for the configuration file,
and open or close the software trigger.
The Plot module is used to select drawing switch, drawing type and digitizer board number
when there are multiple digitizers in the system.
The reading speed of data acquisition and the trigger frequency are exported in the Log module.
There are also the hardware link information, and the error information occurred during the software operation.

The OutFile module is used to set the output file path, file type (ASCII or binary), file name, record time
and the number of files required to be recorded.
There are also automatic target and file compression switches in this module.
The event with no signal will not be recorded in the file when the file compression is opened.
The ControlTarget module is used to set the cycle sequence of the target when the target is automatically changed.
There are five targets available.

The public module is used to show the time of record, connect or
disconnect the waveform digitizer, start or stop acquisition, and exit software.

The software realizes the automatic alarm function of beam loss.
It is the standard for beam loss that there is a signal  of the high frequency trigger of accelerator but
 no signal of the neutron detector at the same time.
Automatic alarm function is automatically opened only when the data file is recording.
\subsection{Control target software and stepper motor}
\begin{figure}
  \centering
  \includegraphics[width=10cm]{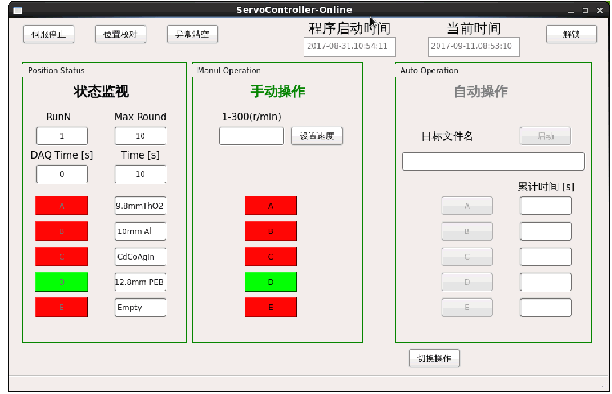}\\
  \caption{Control target software operation interface.}\label{cts}
\end{figure}

\begin{figure}
  \centering
  \includegraphics[width=10cm]{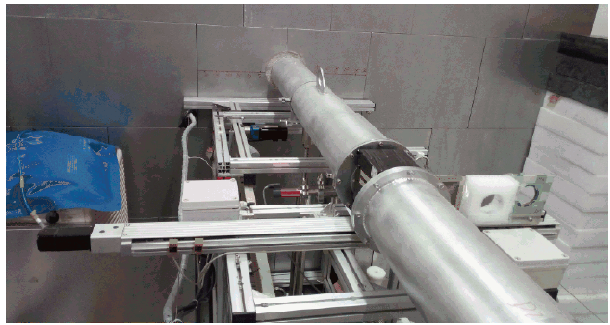}\\
  \caption{Stepper motor and five targets.}\label{motor}
\end{figure}

The control target software is used to control the target position,
 as shown in Fig.~\ref{cts}.
 The software can be operated manually or control the target automatically in combination with
 the data acquisition software.
 The stepper motor and five targets are shown in Fig.~\ref{motor}.

\subsection{Data analysis offline}
\begin{figure}
  \centering
  \includegraphics[width=10cm]{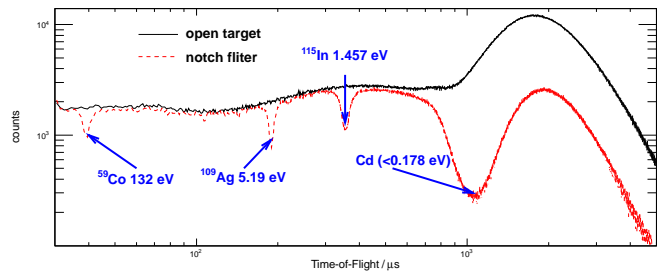}\\
  \caption{Neutron TOF spectrum of the absorber in PNS.}\label{absorber}
\end{figure}

Data analysis offline includes the display of waveform information, pulse amplitude analysis,
pulse area integral, pulse shape discrimination, time of flight spectrum analysis, and multi-pulse analysis, etc..
Neutron is identified using pulse shape discrimination method.
The time of flight (TOF) for neutron is the time difference between the time of arrival of gamma flash pulse
and the time of arrival of neutron pulse in PNS.
Figure~\ref{absorber} is the neutron flight time spectrum of the absorber measured in PNS~\cite{llx1,llx2}.

\section{Conclusion}
\label{cls}
The data acquisition system makes the detector, reaction target and waveform digitizer operation easier and convenient.
Waveform digital technology reduces the number of electronics modules, simplifies the logical circuit, saves costs and improves efficiency.
The system implements automatic timed change targets and files, and automatic alarm, which make experiments easier for operators.
The raw data is compressed and the output file size is reduced.
The analysis offline of waveform data has realized many data analysis functions depending on the purpose of the experiment.

\section*{Acknowledgements}

The authors would like to express their sincere thanks to the staff of Free Electron Group and
other faculty for the excellent operation and strong support.
This work was supported by the Frontier Science Key Program of the Chinese Academy of Sciences (No. QYZDY-SSW-JSC016),
 National Natural Science Foundation of China (NSFC) (No. 11305239 ,  No.91326201£¬ No.11475245)£¬
 and the Chinese TMSR Strategic Pioneer Science and Technology Project (No.XDA02010100).





\end{document}